\documentclass[10pt,twocolumn,showpacs,preprintnumbers,amsmath,amssymb,aps,prb,longbibliography,superscriptaddress]{revtex4-2}
\usepackage{array,braket,mathtools,siunitx,bbm,enumitem}
\usepackage[unicode=true,
 bookmarks=true,bookmarksnumbered=true,bookmarksopen=false,
 breaklinks=true,pdfborder={0 0 0},backref=false,colorlinks=true]{hyperref}
\hypersetup{citecolor={blue},urlcolor={magenta}}
\usepackage{cleveref}
\usepackage{xcolor}
\crefname{appendix}{App.}{Apps.}
\crefname{equation}{Eq.}{Eqs.}
\crefname{figure}{Fig.}{Figs.}
\crefname{table}{Tab.}{Tabs.}
\crefname{section}{Sec.}{Secs.}
\newcommand{\Z}{\mathbb{Z}}
\newcommand{\Complex}{\mathbb{C}}
\newcommand{\Real}{\mathbb{R}}

\begin{document}
\title{Defect Bootstrap: Tight Ground State Bounds in Spontaneous Symmetry Breaking Phases}
\author{Michael G. Scheer}
\affiliation{Department of Physics, Harvard University, Cambridge, Massachusetts 02138, USA}
\author{Nisarg Chadha}
\affiliation{Department of Physics, Harvard University, Cambridge, Massachusetts 02138, USA}
\author{Da-Chuan Lu}
\affiliation{Department of Physics, Harvard University, Cambridge, Massachusetts 02138, USA}
\affiliation{Department of Physics and Center for Theory of Quantum Matter, University of Colorado, Boulder, Colorado 80309, USA}
\author{Eslam Khalaf}
\affiliation{Department of Physics, Harvard University, Cambridge, Massachusetts 02138, USA}
\date{\today}

\begin{abstract}
The recent development of bootstrap methods based on semidefinite relaxations of positivity constraints has enabled rigorous two-sided bounds on local observables directly in the thermodynamic limit. However, these bounds inevitably become loose in symmetry broken phases, where local constraints are insufficient to capture long-range order. In this work, we identify the origin of this looseness as order parameter defects which are difficult to remove using local operators. We introduce a \emph{defect bootstrap} framework that resolves this limitation by embedding the system into an auxiliary \emph{defect model} equipped with ancilla degrees of freedom. This construction effectively enables local operators to remove order parameter defects, yielding tighter bounds in phases with spontaneous symmetry breaking. This approach can be applied broadly to pairwise-interacting local lattice models with discrete or continuous internal symmetries that satisfy a property we call \emph{defect diamagnetism}, which requires that the ground state energy does not decrease upon adding any finite number of symmetry defects. Applying the method to the transverse field Ising models in 1D and 2D, we obtain significantly improved bounds on energy densities and spin correlation functions throughout the symmetry broken phase in 1D and deep within the phase in 2D. Our results demonstrate that physically motivated constraint sets can dramatically enhance the power of bootstrap methods for quantum many-body systems.
\end{abstract}

\maketitle

\begin{figure}
	\centering
	\includegraphics{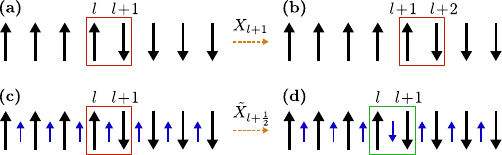}
	\caption{Illustration of perturbative positivity using \textbf{(a)}-\textbf{(b)} $H_{\text{TFIM}}$ in \cref{eq:H-TFIM-1D} and \textbf{(c)}-\textbf{(d)} $H_{\text{defect}}$ in \cref{eq:H-defect-1D} all with $g = 0$. \textbf{(a)} A 1D TFIM state with a domain wall between sites $l$ and $l+1$. \textbf{(b)} Applying $X_{l+1}$ moves the domain wall but does not lower the energy. \textbf{(c)} The same domain wall state, but now with ancilla spins. \textbf{(d)} Applying $\tilde{X}_{l+\frac{1}{2}}$ effectively removes the domain wall and lowers the energy.}
	\label{fig:spins}
\end{figure}

\emph{Introduction}---Characterizing ground states of strongly interacting quantum systems remains one of the central challenges in many-body physics. Although many methods have been developed to approximate ground state observables for quantum many-body Hamiltonians, very few provide rigorous guarantees or certificates, especially for infinite systems (i.e., in the thermodynamic limit). Recently, a class of semidefinite relaxation methods often called \emph{bootstrap} methods have been introduced. The simplest version of bootstrap involves minimizing the energy as a function of certain expectation values, subject to positivity constraints \cite{Mazziotti2020,Mazziotti2023,Barthel2012,Baumgratz2012,Haim2020,Lin2022,Khoo2024,Gao2024,Gao2025,Han2020,Berenstein2021,Berenstein2024,Scheer2024,Schouten2025}. This approach provides a guaranteed lower bound on the ground state energy, making it complementary to variational approaches. It also provides estimates of other ground state expectation values, but these results do not come with rigorous guarantees. Some works have incorporated variational energies to obtain rigorous bounds on ground state expectation values \cite{Han2020a,Wang2024}. However, strong variational upper bounds are not typically available for infinite size systems in dimensions $2$ or higher.

Remarkably, recent works \cite{Fawzi2024,Cho2025} have extended this method to produce rigorous two-sided bounds on local expectation values in local lattice models, both of finite or infinite size, without relying on variational energy bounds. This generalization relies on a ground state specific condition called \emph{perturbative positivity} which formalizes the requirement that the energy cannot be lowered by any operator. Although our focus is on ground states, we note that perturbative positivity can be generalized to select for equilibrium Gibbs or KMS states at nonzero temperature \cite{Fawzi2024, Cho2025}.

The ability to obtain rigorous two-sided bounds directly in the thermodynamic limit sets this approach apart from most other numerical methods and has the potential to resolve many outstanding questions about ground state physics. However, to achieve this potential, the bounds need to be sufficiently tight to provide meaningful restrictions on the expectation values of observables \footnote{We say that a two-sided bound is tight if the difference between the upper and lower limits is small. If the difference is large, we say that the bound is loose even if the true value lies close to one of the limits.}. The tightness of the bounds depends strongly on the choice of operators used to define the constraints, and the bounds can be quite loose even in simple models.

A relevant example can be found in Ref. \cite{Fawzi2024a} which applied bootstrap to one of the simplest lattice spin models, the 1D transverse field Ising model (TFIM)
\begin{equation}
H_{\text{TFIM}} = -\sum_j Z_j Z_{j+1} - g \sum_j X_j.
\end{equation}
Here, $X_j$, $Y_j$, and $Z_j$ are Pauli operators for site $j$ and $g$ is the transverse field. The Hamiltonian has a global $\Z_2$ spin-flip symmetry, $\hat{X} H_{\text{TFIM}} \hat{X}^\dagger = H_{\text{TFIM}}$ where $\hat{X} = \prod_j X_j$. For $g = 0$ there are two ground states with all spins aligned in the $+z$ or the $-z$ direction, while for $g \to \infty$, there is a unique ground state with all spins pointing in the $+x$ direction. There is a spontaneous symmetry breaking (SSB) phase transition at the critical field $g = 1$ \cite{Sachdev2011}. By using all local operators with support on a $5$ site region, Ref. \cite{Fawzi2024a} obtained bounds in the thermodynamic limit which are tight in the symmetric phase $g > 1$ but are quite loose in the SSB phase $g < 1$. Ref. \cite{Cho2025a} used a sophisticated coarse-graining approach to push the size of the region to $20$, but the obtained bounds were still loose in the SSB phase. This suggests  a fundamental limitation in obtaining tight bounds using only local operator constraints in this regime.

This issue can be understood with a simple heuristic argument. In the symmetry breaking phase $g < 1$, states with large domains have low energy. Domain walls, which are the boundaries between domains, are defects in the order parameter that cannot be created or destroyed by any local operator. Instead, one can only annihilate them in pairs by flipping entire domains. Since perturbative positivity works by ruling out states whose energy can be lowered by the application of an operator in the constraint set, the domain walls are an obstruction to tight bootstrap bounds.

We show that this problem can be solved by bootstrapping a \emph{defect model} containing ancilla spins on the edges which effectively allow for the removal of domain walls using local operators. We prove that ground state bounds for the TFIM can be evaluated directly within the defect model. Additionally, we show that this approach is generally applicable to any pairwise-interacting local lattice model with a discrete or continuous internal symmetry provided that it satisfies a property we call \emph{defect diamagnetism}. Roughly speaking, an infinite size model has defect diamagnetism if the ground state energy does not decrease upon adding any finite number of symmetry defects. This is a natural condition, and for infinite fermionic models with $U(1)$ symmetry, it is essentially equivalent to orbital diamagnetism.

We apply this \emph{defect bootstrap} method to the infinite 1D and 2D TFIMs. In 1D, we find that bounds in the SSB phase are significantly tigher with the defect model. Specifically, we obtain tight bounds on the energy density and on spin correlators $\braket{Z_0 Z_r}$ away from the critical point $g = 1$. In 2D, we find significant improvement deep inside the SSB phase $0 \leq g \lesssim 1$, leading to tight bounds everywhere except in the region $1 \lesssim g \lesssim 3$ (the critical point is $g \approx 3.04$ \cite{Liu2013}). Since the 2D TFIM has not been exactly solved for $g > 0$, these represent, to the best of our knowledge, the first non-trivial two-sided bounds on ground state spin correlations in the thermodynamic limit. Our work establishes defect bootstrap as a powerful tool to obtain rigorous bounds on the properties of SSB phases and highlights the importance of physically-motivated constraint operators.

\emph{Bootstrap for finite systems}---We begin by reviewing the bootstrap method for finite size systems. We will allow the possibility that the Hilbert space is infinite dimensional, as this will make the generalization to infinite systems more clear. In the case of an infinite dimensional Hilbert space, the trace of an operator is not always well defined, for example if it is unbounded. For that reason, we will be careful to specify the analytic properties required of various operators.

Let $\mathcal{B}$ be the algebra of bounded operators and let the Hamiltonian $H \in \mathcal{B}$ be a bounded Hermitian operator. One can prove that a trace class operator $\rho$ is a ground state density operator for $H$ if and only if the following three conditions are satisfied:
\begin{enumerate}[itemsep=-0.3cm]
\item (Normalization) $\text{tr}(\rho) = 1$.\\
\item (Positivity) $\text{tr}(\mathcal{O}^\dagger\mathcal{O}\rho) \geq 0$ for all $\mathcal{O} \in \mathcal{B}$.\\
\item (Perturbative positivity) $\text{tr}(\mathcal{O}^\dagger [H, \mathcal{O}] \rho) \geq 0$ for all $\mathcal{O} \in \mathcal{B}$.
\end{enumerate}
The normalization and positivity conditions together are equivalent to saying that $\rho$ represents a physical state. The perturbative positivity condition morally says that no $\mathcal{O} \in \mathcal{B}$ can lower the energy of $\rho$, and it is a property satisfied only by ground states \cite{Fawzi2024,Cho2025}.

\begin{figure}
	\centering
	\includegraphics{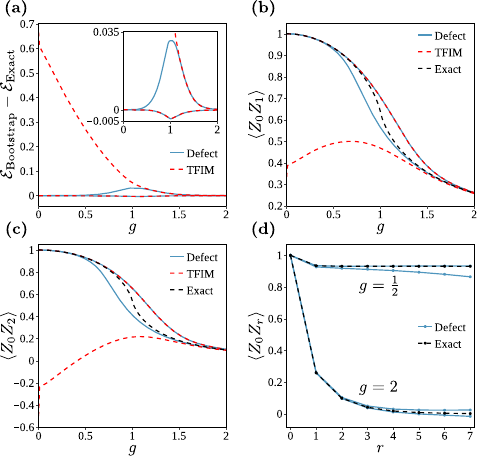}
	\caption{Bootstrap bounds for the infinite 1D TFIM using the Hamiltonians $H_{\text{TFIM}}$ (red dashed lines) and $H_{\text{defect}}$ (blue solid lines) in \cref{eq:H-TFIM-1D,eq:H-defect-1D} with $L = \infty$. The black dashed lines in \textbf{(b)}-\textbf{(d)} show the exact values. \textbf{(a)} The energy density relative to the exact energy density. The inset shows a zoomed in plot. \textbf{(b)}-\textbf{(c)} The spin correlator $\braket{Z_0 Z_r}$ for $r = 1, 2$ as a function of $g$. \textbf{(d)} The spin correlator $\braket{Z_0 Z_r}$ as a function of $r$ for $g = \frac{1}{2}, 2$ using $H_{\text{defect}}$. The corresponding results with $H_{\text{TFIM}}$ are shown in \cref{fig:1D_2D}.}
	\label{fig:1D}
\end{figure}

Now let $\mathcal{P} \subset \mathcal{B}$ be a finite set of bounded operators and let $R \in \mathcal{B}$ be a bounded Hermitian operator. We define $\braket{R}^{\min}_{H,\mathcal{P}}$ by the optimization problem
\begin{equation}\label{eq:R-P-min}
\begin{split}
\braket{R}^{\min}_{H,\mathcal{P}} = &\min_\rho \text{tr}(R \rho)\\
&\text{s.t. } \text{tr}(\rho) = 1,\\
&\text{tr}(\mathcal{O}^\dagger \mathcal{O} \rho) \geq 0 \text{ for all } \mathcal{O} \in \text{Span}(\mathcal{P}),\\
&\text{tr}(\mathcal{O}^\dagger[H, \mathcal{O}] \rho) \geq 0 \text{ for all } \mathcal{O} \in \text{Span}(\mathcal{P})
\end{split}
\end{equation}
and additionally define $\braket{R}^{\max}_{H,\mathcal{P}} = -\braket{-R}^{\min}_{H,\mathcal{P}}$. Since we have relaxed the positivity and perturbative positivity conditions, we have
\begin{equation}\label{eq:bounds}
\braket{R}^{\min}_{H,\mathcal{P}} \leq \text{tr}(R \rho) \leq \braket{R}^{\max}_{H,\mathcal{P}}
\end{equation}
for any ground state density operator $\rho$. We now introduce a finite linearly independent set $\mathcal{Q} \subset \mathcal{B}$ of bounded Hermitian operators such that $I, R \in \text{Span}(\mathcal{Q})$ where $I$ is the identity operator, and $p_1^\dagger p_2, p_1^\dagger [H, p_2] \in \text{Span}(\mathcal{Q})$ for all $p_1, p_2 \in \mathcal{P}$. We now see that \cref{eq:R-P-min} only depends on $\rho$ through the finite collection of expectation values $\text{tr}(q\rho)$ for $q \in \mathcal{Q}$. By arguments very similar to those in Ref. \cite{Scheer2024}, when \cref{eq:R-P-min} is thought of as a minimization over the real variables $\text{tr}(q\rho)$ for $q \in \mathcal{Q}$, it becomes a semidefinite program (SDP) \cite{Vandenberghe1996,Boyd2004}, which is a type of convex optimization problem that can be solved using resources polynomial in the size of $\mathcal{P}$ \footnote{We use MOSEK \cite{Mosek} to numerically solve SDPs in this work.}.

We now consider the role of symmetry. We say that a unitary or anti-unitary operator $U$ is a symmetry of $H$ if $U H U^\dagger = H$. If $H$ has a group $G$ of symmetries, then one can always find a ground state density operator $\rho$ satisfying $U \rho U^\dagger = \rho$ for all $U \in G$ \footnote{Sometimes this requires $\rho$ to be a mixed state rather than a pure state.}. We can then include the symmetry constraints
\begin{equation}\label{eq:symmetry-constraints}
\text{tr}(q_1 \rho) = \text{tr}(q_2 \rho) \text{ whenever } q_1 = U q_2 U^\dagger
\end{equation}
for all $q_1, q_2 \in \text{Span}(\mathcal{Q})$ and $U \in G$ in the optimization problem in \cref{eq:R-P-min}. The symmetry constraints in \cref{eq:symmetry-constraints} can be used to significantly reduce the size of the SDP using the methods described in Ref. \cite{Scheer2024}.

For a symmetry $U \in G$ that satisfies $U R U^\dagger = R$, the inclusion of its symmetry constraint will not change the values of the bounds $\braket{R}^{\min}_{H, \mathcal{P}}$ and $\braket{R}^{\max}_{H, \mathcal{P}}$. Therefore if $R$ is invariant under all $U \in G$ then the bounds obtained with all constraints in \cref{eq:symmetry-constraints} satisfy \cref{eq:bounds} for any ground state $\rho$, whether or not it is symmetric.

On the other hand, if $U R U^\dagger \neq R$, the inclusion of its symmetry constraint may change the bounds \footnote{An example in the 1D TFIM concerns the symmetry $U = \hat{X}$. If we take $R = Z_0 Z_1$ then since $U R U^\dagger = R$, the symmetry constraints will not change the bounds. However if $R = Z_0$ then $U R U^\dagger = -R$, and in this case $U \rho U^\dagger = \rho$ implies $\text{tr}(\rho R) = 0$.}. As a result, in general the bounds obtained with all constraints in \cref{eq:symmetry-constraints} imposed only satisfy \cref{eq:bounds} for ground states $\rho$ which are invariant under all $U \in G$ for which $U R U^\dagger \neq R$.

Finally, we consider the case in which one knows that the Hamiltonian $H$ has multiple ground states. In this case, it may be desirable to select a particular ground state using linear constraints. If $\mathcal{C} \subset \mathcal{B}$ is a finite set of bounded Hermitian operators, then we can include
\begin{equation}\label{eq:hermitian-constraints}
\text{tr}(C \rho) = 0 \text{ for all } C \in \mathcal{C}
\end{equation}
in the optimization problem in \cref{eq:R-P-min}, provided that there exists a ground state $\rho$ satisfying \cref{eq:hermitian-constraints}. Note that we must have $I \not\in \text{Span}(\mathcal{C})$ so as to not contradict $\text{tr}(\rho) = 1$. If symmetry constraints of the form in \cref{eq:symmetry-constraints} for $U \in G$ are included, we must have
\begin{equation}
I \not\in \text{Span}(\{U C U^\dagger | U \in G, C \in \mathcal{C}\}).
\end{equation}

\begin{figure}
	\centering
	\includegraphics{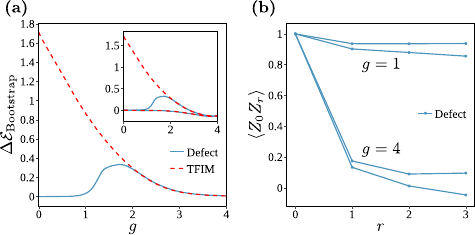}
	\caption{Bootstrap results for the infinite 2D TFIM using the Hamiltonians $H_{\text{TFIM}}$ (red dashed lines) and $H_{\text{defect}}$ (blue solid lines) in \cref{eq:general-H-TFIM,eq:general-H-defect} with $L = \infty$. \textbf{(a)} The difference between the upper and lower bootstrap bounds for the energy density. The inset shows the bounds for the energy density relative to the mean field value $\mathcal{E}_{\text{Variational}}$ in \cref{eq:ansatz} with $d = 2$. \textbf{(b)} The spin correlator $\braket{Z_0 Z_r}$ as a function of $r$ for $g = 1, 4$ using $H_{\text{defect}}$. The corresponding results with $H_{\text{TFIM}}$ are shown in \cref{fig:1D_2D}.}
	\label{fig:2D}
\end{figure}

\emph{Bootstrap for infinite systems}---The generalization of bootstrap to infinite systems must confront a few conceptual difficulties. First, the Hamiltonian may not be a well-defined operator since the energy is an extensive quantity. For example, a translation-invariant Hamiltonian will generally have a nonzero ground state energy density, which corresponds to an infinite ground state energy. Similarly, the ground state density matrix may not be a well-defined operator, and even worse there may not even be a well-defined Hilbert space. Thankfully, for infinite systems possessing a notion of locality, these issues can all be resolved using the formalism of algebraic quantum mechanics \cite{Bratteli1987,Bratteli1997}.

Before providing a rigorous formulation, we will explain the rough idea. Regarding the Hamiltonian, the important observation is that $H$ only enters \cref{eq:R-P-min} through its commutators $[H, \mathcal{O}]$ with operators $\mathcal{O} \in \mathcal{P}$. If we choose $\mathcal{P}$ to consist only of local operators, it is sufficient to define $H$ by specifying its commutators with local operators. Similarly, we note that $\rho$ only enters \cref{eq:R-P-min} through its expectation values $\text{tr}(q\rho)$ for $q \in \mathcal{Q}$. If we choose $\mathcal{Q}$ to consist only of local operators, it is sufficient to define $\rho$ by specifying its traces against local operators.

The general formulation using algebraic quantum mechanics proceeds as follows. The algebra $\mathcal{B}$ of bounded operators is generalized to the algebra $\mathcal{A}$ of quasi-local observables, trace class operators $\rho$ are generalized to linear maps $\omega : \mathcal{A}\to \Complex$, the Hamiltonian $H$ is generalized to a derivation map $h : \mathcal{A} \to \mathcal{A}$, and symmetry operators $U$ are generalized to automorphisms $u : \mathcal{A}\to \mathcal{A}$. Under the dictionary
\begin{equation}
\begin{split}
& \mathcal{B} \mapsto \mathcal{A}, \quad \text{tr}(q\rho) \mapsto \omega(q), \quad [H, \mathcal{O}] \mapsto h(\mathcal{O}),\\
& U q U^\dagger \mapsto u(q),\quad U \rho U^\dagger \mapsto \omega \circ u^{-1},\\
& U H U^\dagger \mapsto u \circ h \circ u^{-1},
\end{split}
\end{equation}
bootstrap is generalized to all systems that can be modeled with algebraic quantum mechanics. This includes the infinite lattice models of locally interacting spins we study in this paper.

\emph{Defect model}---We now describe the defect model construction. We begin by considering a 1D TFIM on a finite length $L$ chain with open boundary conditions and Hamiltonian
\begin{equation}\label{eq:H-TFIM-1D}
H_{\text{TFIM}} = -\sum_{j=1}^{L-1} Z_j Z_{j+1} - g\sum_{j=1}^L X_j.
\end{equation}
If we conjugate $H_{\text{TFIM}}$ by $\prod_{j=1}^l X_j$ for some $l \leq L-1$, the resulting Hamiltonian is identical to $H_{\text{TFIM}}$ except that the sign of the term $Z_l Z_{l+1}$ is changed from $-$ to $+$. This $+$ sign induced by a partial application of the $\Z_2$ spin-flip symmetry is called a \emph{defect} \cite{Aasen2016,Sahand2024,Choi2025}. By applying a sequence of unitary conjugations, it is possible to create any of the $2^{L-1}$ defect configurations. We now collect all of these configurations in a single system by introducing ancilla spins with Pauli operators $\tilde{X}_{j+\frac{1}{2}}$, $\tilde{Y}_{j+\frac{1}{2}}$, $\tilde{Z}_{j+\frac{1}{2}}$ for each $1 \leq j \leq L-1$ and defining the defect Hamiltonian \footnote{We note that moving to the defect Hamiltonian is similar to gauging the $\Z_2$ symmetry. The difference is that we do not wish to include a Maxwell term for the gauge field or impose Gauss's law.}
\begin{equation}\label{eq:H-defect-1D}
H_{\text{defect}} = -\sum_{j=1}^{L-1} Z_j \tilde{Z}_{j+\frac{1}{2}} Z_{j+1} - g\sum_{j=1}^L X_j.
\end{equation}
Each sector of $H_{\text{defect}}$ corresponding to definite $1$ or $-1$ values for the $\tilde{Z}$ operators is a defect configuration, and we identify $H_{\text{TFIM}}$ with the sector in which all $\tilde{Z}$ operators are set to $1$. Our preceding argument shows that in fact all $\tilde{Z}$ sectors are unitarily equivalent, so that any ground state of $H_{\text{TFIM}}$ is also a ground state of $H_{\text{defect}}$. This means that bootstrap bounds on ground state expectation values in $H_{\text{defect}}$ apply also to $H_{\text{TFIM}}$. Since this result holds for all finite sizes, it is also true for the infinite 1D TFIM.

We now compare bootstrap programs for $H_{\text{TFIM}}$ and $H_{\text{defect}}$ at $g = 0$ and length $L$. Consider a pure state for $H_{\text{TFIM}}$ in which all spins from $1$ to $l$ are in the $+z$ direction and all spins from $l+1$ to $L$ are in the $-z$ direction, as illustrated in \cref{fig:spins}\textbf{(a)}. This state has a single domain wall between sites $l$ and $l+1$ and therefore has energy $3 - L$ which is greater than the ground state energy $1-L$. Furthermore, if we apply $X_l$ or $X_{l+1}$ to this state, the domain wall simply moves by one site and the energy does not change (see \cref{fig:spins}\textbf{(b)}). This means that perturbative positivity with an operator set $\mathcal{P}$ consisting of $X_j$ operators cannot tell that this state is not a ground state. Next, we consider this same state but in the context of $H_{\text{defect}}$ by taking all ancilla spins in the $+z$ direction (see \cref{fig:spins}\textbf{(c)}). The energy of this state is still $3-L$, but if we apply the $\tilde{X}_{l+\frac{1}{2}}$ operator the energy lowers to $1-L$, the ground state energy (see \cref{fig:spins}\textbf{(d)}). Therefore perturbative positivity with an operator set $\mathcal{P}$ consisting of $\tilde{X}_{j+\frac{1}{2}}$ operators can tell that this state is not a ground state. The general point is that when using $H_{\text{defect}}$, a domain wall can be effectively removed by the introduction of a defect.

We now consider the generalization to a 2D TFIM on a finite $L \times L$ square lattice with open boundary conditions and Hamiltonian
\begin{equation}\label{eq:general-H-TFIM}
H_{\text{TFIM}} = -\sum_{\braket{v,w}} Z_v Z_w - g\sum_v X_v
\end{equation}
where $\braket{v,w}$ means that sites $v$ and $w$ are neighbors. We again introduce ancilla spins with Pauli operators $\tilde{X}_{v,w}, \tilde{Y}_{v,w}, \tilde{Z}_{v,w}$ for all neighbors $v$ and $w$ and define the defect Hamiltonian
\begin{equation}\label{eq:general-H-defect}
H_{\text{defect}} = -\sum_{\braket{v,w}} Z_v \tilde{Z}_{v,w} Z_w - g \sum_v X_v.
\end{equation}
As before, $H_{\text{defect}}$ contains all defect configurations as $\tilde{Z}$ sectors. The important difference from the 1D case is that these configurations are no longer all equivalent up to conjugation by products of $X$ operators. To see this, note that when $H_{\text{TFIM}}$ is conjugated by $X_v$ for some site $v$, the resulting Hamiltonian is identical to $H_{\text{TFIM}}$ except that the signs of the $Z_v Z_w$ terms for neighbors $w$ of $v$ have changed from $-$ to $+$. Since the product of the signs around any closed loop is invariant under this process, it is not possible to achieve a configuration with a single defect. Nonetheless, we prove in \cref{app:trivial-sectors,app:TFIM} that all ground states of $H_{\text{TFIM}}$ are ground states of $H_{\text{defect}}$, and as a result bootstrap bounds on ground state expectation values in $H_{\text{defect}}$ apply also to $H_{\text{TFIM}}$ at both finite and infinite size.

Finally, we discuss the general construction of the defect model. As shown in \cref{app:general-defect}, it is possible to define a defect model $H_{\text{defect}}$ for any pairwise-interacting local lattice Hamiltonian $H$ with an internal symmetry group $G$. For each pair of interacting degrees of freedom in $H$, the defect model includes an ancilla Hilbert space with basis states corresponding to the elements of $G$, and $H$ can be identified with the \emph{identity sector} of $H_{\text{defect}}$, by which we mean the sector in which all ancilla states are the identity.

Unlike the case of the TFIM, generally one may find that ground states of $H$ are not ground states of $H_{\text{defect}}$. Suppose for now that $H$ is a finite system, and let $E_0$ be its ground state energy. If for some constraint set $\mathcal{P}$ one has
\begin{equation}\label{eq:defect-diamagnetism}
\braket{\psi | \mathcal{O}^\dagger H_{\text{defect}} \mathcal{O} | \psi} \geq E_0 \braket{\psi | \mathcal{O}^\dagger \mathcal{O} | \psi} \text{ for all } \mathcal{O} \in \mathcal{P}
\end{equation}
for all pure states $\ket{\psi}$ in the identity sector, then any ground state $\rho$ for $H$ satisfies perturbative positivity with respect to $H_{\text{defect}}$ and $\mathcal{P}$, despite the fact that it may not be a ground state for $H_{\text{defect}}$. It follows that we can use \cref{eq:R-P-min} with $H$ replaced by $H_{\text{defect}}$ to obtain rigorous bounds on ground state expectation values for $H$. In order to obtain tight bounds, one will typically need to include \cref{eq:hermitian-constraints} where the constraint operators $C$ restrict $\rho$ to the identity sector. When \cref{eq:defect-diamagnetism} holds, we say that $H$ has \emph{defect diamagnetism} with respect to $\mathcal{P}$.

Now suppose that $H$ is an infinite size system which can be realized as a limit of finite size systems $H = \lim_{n\to\infty} H^{(n)}$ with ground state energies $E^{(n)}_0$. Let $H^{(n)}_{\text{defect}}$ and $I^{(n)}$ denote the defect Hamiltonian for $H^{(n)}$ and its identity sector, respectively. If for every local operator $\mathcal{O}$ we have \footnote{In this expression, it is possible that $\mathcal{O}\ket{\psi} = 0$ in which case the numerator and denominator both vanish. In this case, we regard the ratio as $+\infty$.}
\begin{equation}\label{eq:defect-diagmagnetism-infinite}
\lim_{n\to\infty} \min_{\ket{\psi} \in I^{(n)}} \frac{\braket{\psi | \mathcal{O}^\dagger H^{(n)}_{\text{defect}} \mathcal{O} | \psi}}{\braket{\psi | \mathcal{O}^\dagger \mathcal{O} | \psi}} - E^{(n)}_0 \geq 0
\end{equation}
we say that $H$ has defect diamagnetism. In this case, rigorous bounds on ground state expectation values for $H$ can be obtained by bootstrapping with $H_{\text{defect}}$. It is important to note that while the definition of defect diamagnetism for a finite system depends on the constraint set $\mathcal{P}$, the definition for an infinite system does not.

As an example, consider an infinite 2D local lattice model of fermions with particle number $U(1)$ symmetry. In this case, moving to the defect model is equivalent to the Peierls substitution, so the sectors of $H_{\text{defect}}$ correspond to different applied magnetic field configurations. In this case, \cref{eq:defect-diagmagnetism-infinite} simply says that adding a local magnetic field cannot lower the ground state energy. This property will certainly hold for systems with orbital diamagnetism.

\emph{Numerical results}---\cref{fig:1D} shows bootstrap results for the infinite 1D TFIM using both $H_{\text{TFIM}}$ in \cref{eq:H-TFIM-1D} and $H_{\text{defect}}$ in \cref{eq:H-defect-1D}. In both cases, we solve \cref{eq:R-P-min} for various operators $R$, making use of translation, inversion, and spin-flip symmetries with \cref{eq:symmetry-constraints}. For $H_{\text{defect}}$, we include \cref{eq:hermitian-constraints} with $\mathcal{C} = \{(\tilde{Z}_{\frac{1}{2}} - I)^2\}$ which requires $\rho$ to be in the sector corresponding to $H_{\text{TFIM}}$ \footnote{Since we are including translation symmetry, it suffices to include the linear constraint at only one site.}, and we additionally include the symmetry $\tilde{Z}_{j+\frac{1}{2}}$ for each integer $j$. The constraint sets $\mathcal{P}$ are described in \cref{app:constraint-sets}.

\cref{fig:1D}\textbf{(a)} shows the energy density relative to the exact value as a function of transverse field for both Hamiltonians \footnote{The exact formulas for ground state correlations in the 1D TFIM can be found in Ref. \cite{Sachdev2011}.}. As in Refs. \cite{Fawzi2024a,Cho2025a}, the energy density lower bound for $H_{\text{TFIM}}$ is tight everywhere, but the energy density upper bound is tight only in the symmetry preserving regime $g > 1$. In contrast, the energy density lower and upper bounds for $H_{\text{defect}}$ are tight everywhere (even at the critical point $g = 1$, the error is less than $0.035$). \cref{fig:1D}\textbf{(b)} and \textbf{(c)} show bounds for the spin correlators $\braket{Z_0 Z_1}$ and $\braket{Z_0 Z_2}$. The results are similar to the case of energy density, except that the $H_{\text{defect}}$ bounds are not quite as tight near the critical point $g = 1$. \cref{fig:1D}\textbf{(d)} shows bounds on $\braket{Z_0 Z_r}$ as a function of $r$ for $H_{\text{defect}}$ and $g = \frac{1}{2}, 2$. For $g = \frac{1}{2} < 1$ we see long-range spin correlations, while for $g = 2 > 1$ the correlations decay to $0$. The corresponding results for $H_{\text{TFIM}}$ are shown in \cref{fig:1D_2D}\textbf{(a)} and \textbf{(b)}. Here, the lower bound on $\braket{Z_0 Z_r}$ for $g = \frac{1}{2}$ is very loose. While it is true that the ancilla degrees of freedom in the defect model generally increase the size of the SDPs we must solve, we show in \cref{fig:1D_2D_big} that simply increasing the size of the constraint set $\mathcal{P}$ for $H_{\text{TFIM}}$ is not able to significantly tighten the bounds.

\cref{fig:2D} shows bootstrap results for the infinite 2D TFIM using both $H_{\text{TFIM}}$ in \cref{eq:general-H-TFIM} and $H_{\text{defect}}$ in \cref{eq:general-H-defect}. We again solve \cref{eq:R-P-min}, this time making use of translation, rotation, reflection, and spin-flip symmetries with \cref{eq:symmetry-constraints}. For $H_{\text{defect}}$, we include \cref{eq:hermitian-constraints} with $\mathcal{C} = \{(\tilde{Z}_{(\frac{1}{2},\frac{1}{2})} - I)^2\}$ and symmetries $\tilde{Z}_{v,w}$ for each pair of neighbors $v, w$. The constraint sets $\mathcal{P}$ are described in \cref{app:constraint-sets}.

\cref{fig:2D}\textbf{(a)} shows the difference between the upper and lower bounds for the energy density. Additionally, the inset shows the bounds relative to the simple mean-field ansatz derived in \cref{app:ansatz}. For both Hamiltonians, the energy density lower bound matches the ansatz at $g = 0$, and falls slightly below the ansatz for large $g$, which demonstrates that the lower bounds are accurate. On the other hand, for $H_{\text{TFIM}}$ the upper bound is much higher than the ansatz for $0 \leq g \lesssim 3$ while for $H_{\text{defect}}$ the upper bound matches the ansatz for small $g$ but is much higher that the ansatz for $1 \lesssim g \lesssim 3$. We conclude that moving to the defect model significantly improves the energy upper bound deep in the symmetry broken regime $0 \leq g \lesssim 1$, but not in the intermediate regime $1 \lesssim g \lesssim 3$.

\cref{fig:2D}\textbf{(b)} shows bounds on $\braket{Z_0 Z_r}$ as a function of $r$ for $H_{\text{defect}}$ and $g = 1,4$. The results suggest that $g = 1$ ($g = 4$) lies in the symmetry breaking (symmetry preserving) phase, which is consistent with the critical field $g \approx 3.04$ from Ref. \cite{Liu2013}. The corresponding results for $H_{\text{TFIM}}$ are shown in \cref{fig:1D_2D}\textbf{(c)} and \textbf{(d)}. Here, the lower bound on $\braket{Z_0 Z_r}$ for $g = 1$ is very loose.

\emph{Discussion}---Refs. \cite{Fawzi2024a,Cho2025a} both succeeded in bootstrapping reasonably accurate bounds for the symmetry breaking order parameter $\braket{Z_0}$ for the infinite 1D TFIM. These bounds show that $\braket{Z_0}$ is very close to $0$ in the symmetry preserving phase $g > 1$, but for $g \to 0$ they inevitably become loose since any value of $\braket{Z_0}$ in the interval $[-1, 1]$ is actually allowed. More generally, one can provide strong evidence against a given SSB order in some model by bounding the order parameter expectation value to a small region around $0$. The more difficult task is to provide evidence that SSB is actually present, which involves showing that the order parameter has long-range correlations. The source of the difficulty is that order parameter defects such as domain walls in the case of the TFIM are obstructions to the tightness of bootstrap bounds using local constraint operators. We have shown in the case of the 1D and 2D TFIMs that the defect model construction produces non-trivial lower bounds on long-range order parameter correlations, at least deep in the SSB phase. We believe this will also hold for many other models with SSB, but we leave this for future work.

\emph{Acknowledgements}---We thank Pietro M. Bonetti, Jonah Herzog-Arbeitman, and Ashvin Vishwanath for helpful discussions. N.C. is supported by NSF award No. 2441781. D.C.L. is supported by the Simons Collaboration on Ultra-Quantum Matter, which is a grant from the Simons Foundation (grant No. 651440).

\bibliography{bibliography}
\appendix

\begin{figure}
	\centering
	\includegraphics{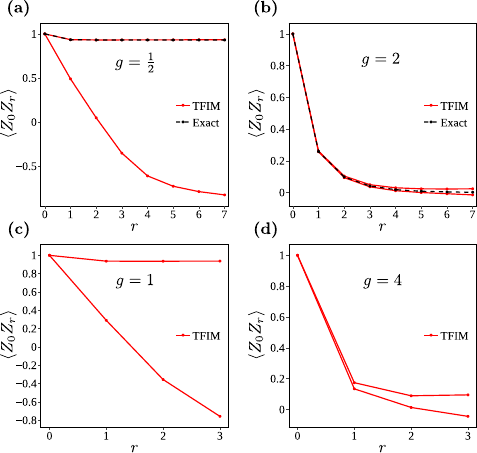}
	\caption{Bootstrap bounds on spin correlators $\braket{Z_0 Z_r}$ for infinite TFIMs without using the defect model. \textbf{(a)}-\textbf{(b)} 1D TFIM with Hamiltonian $H_{\text{TFIM}}$ in \cref{eq:H-TFIM-1D} with $L = \infty$ and $g = \frac{1}{2}, 2$. The red solid lines show the bootstrap bounds while the black dashed lines show the exact value. \textbf{(c)}-\textbf{(d)} 2D TFIM with Hamiltonian $H_{\text{TFIM}}$ in \cref{eq:general-H-TFIM} with $L = \infty$ and $g = 1,4$. Corresponding results for $H_{\text{defect}}$ are shown in \cref{fig:1D}\textbf{(d)} and \cref{fig:2D}\textbf{(b)}.}
	\label{fig:1D_2D}
\end{figure}

\section{Constraint sets}\label{app:constraint-sets}
We now describe the constraint sets $\mathcal{P}$ used in the calculations for \cref{fig:1D,fig:2D,fig:1D_2D}. For a square lattice TFIM in $d$ spatial dimensions, the spins are labeled by vectors in $\Z^d$. For the defect model, the ancilla spins are labeled by vectors of the form $v + \frac{e_j}{2}$ where $v \in \Z^d$ and $e_j$ is the $j$th standard basis vector. We define distances between spins using the taxicab metric $m(v) = \sum_{j=1}^d |v_j|$.

In all cases, the set $\mathcal{P}$ consists of products of Pauli operators for spins in the model. For each product, we define the ``distance" to be the maximal taxicab metric of any constituent Pauli operator. Additionally, the ``diameter" is the maximal taxicab distance between Pauli operators, and the ``degree" is the total number of Pauli operators.

For both models in 1D, the set $\mathcal{P}$ consists of all products with distance at most $5$, diameter at most $1$, and degree at most $3$. For $H_{\text{TFIM}}$ ($H_{\text{defect}}$) the number of elements of $\mathcal{P}$ is $124$ ($928$). For both models in 2D, the set $\mathcal{P}$ consists of all products with distance at most $2$, diameter at most $1$, and degree at most $3$. For $H_{\text{TFIM}}$ ($H_{\text{defect}}$) the number of elements of $\mathcal{P}$ is $184$ ($3220$).

\begin{figure}
	\centering
	\includegraphics{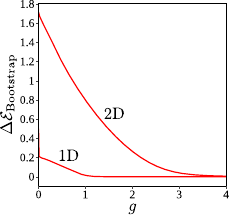}
	\caption{Bootstrap results for the infinite 1D and 2D TFIMs without using the defect model, but with larger constraint sets $\mathcal{P}$ than used in \cref{fig:1D,fig:2D,fig:1D_2D} (see \cref{app:constraint-sets} for details). Specifically, we show the difference between the energy density upper and lower bounds as a function of transverse field. These bounds are loose in the SSB phase despite the increased number of constraint operators.}
	\label{fig:1D_2D_big}
\end{figure}

In order to demonstrate that the improved bounds from the defect model are not simply the result of using an increased number of constraints, we show in \cref{fig:1D_2D_big} the difference between energy density upper and lower bounds for $H_{\text{TFIM}}$ in both 1D and 2D using larger constraint sets $\mathcal{P}$. The 1D result uses the set $\mathcal{P}$ consisting of all products with distance at most $5$, diameter at most $2$, and degree at most $3$. This set contains $448$ elements. The 2D result uses the set $\mathcal{P}$ consisting of all products with distance at most $2$ and either diameter at most $1$ and degree at most $3$ or diameter at most $2$ and degree at most $2$. This set contains $418$ elements. In both cases, the bootstrap bounds using these larger constraint sets are still quite loose in the SSB phases.

\section{Mean field ansatz}\label{app:ansatz}
Consider the infinite square lattice TFIM in $d$ dimensions. A simple ansatz consists of a product state in which the density matrix for each site $v$ is $\rho_v = (I + Z_v \cos\theta + X_v \sin\theta)/2$ for an angle $\theta$ which does not depend on $v$. The energy density of this state is $E_\theta = -d \cos^2\theta - g\sin\theta$. Minimizing over $\theta$ produces the variational energy density upper bound
\begin{equation}\label{eq:ansatz}
\mathcal{E}_{\text{Variational}} = \begin{cases}
-d -\frac{g^2}{4d} & \text{for } 0 \leq g \leq 2d\\
-g & \text{for } 2d \leq g.
\end{cases}
\end{equation}

\section{General defect model construction}\label{app:general-defect}
We now generalize the defect model construction to a broad class of lattice models. We consider a system defined with respect to a graph with vertices $\mathcal{V}$. For each $v \in \mathcal{V}$, let $\mathcal{N}(v)$ be the set of neighbors of $v$, which satisfies
\begin{enumerate}[itemsep=0cm]
\item $v \not\in \mathcal{N}(v)$
\item $w \in \mathcal{N}(v)$ if and only if $v \in \mathcal{N}(w)$.
\end{enumerate}
We consider a Hamiltonian of the form
\begin{equation}\label{eq:general-Hamiltonian}
H = \sum_{v \in \mathcal{V}} \mathcal{H}_v^T \mathcal{S}_v + \sum_{w \in \mathcal{N}(v)} \mathcal{S}_v^T \mathcal{H}_{v,w} \mathcal{S}_w,
\end{equation}
where $\mathcal{S}_v$ is a finite vector of Hermitian operators, $\mathcal{H}_v$ is a real vector, and $\mathcal{H}_{v,w}$ is a complex matrix with $\mathcal{H}_{v,w}^\dagger = \mathcal{H}_{w,v}$. We note that models which are typically written in terms of non-Hermitian operators can still be cast in this form. For example, a fermionic system written in terms of creation and annihilation operators can be rewritten using the Majorana basis. In fact, \cref{eq:general-Hamiltonian} is the general form of a pairwise-interacting local lattice Hamiltonian.

Now suppose a symmetry group $G$ has a unitary representation $U_v$ for each vertex $v \in \mathcal{V}$, which satisfies
\begin{equation}\label{eq:U-condition-1}
U_v(a) U_w(b) = U_w(b) U_v(a) \text{ if } v \neq w
\end{equation}
for all $a, b \in G$, and
\begin{equation}\label{eq:U-condition-2}
U_v(a) \mathcal{S}_w U_v^\dagger(a) = \begin{cases}
\mathcal{S}_w & \text{if } v \neq w\\
V(a^{-1}) \mathcal{S}_w & \text{if } v = w,
\end{cases}
\end{equation}
where $V$ is a real orthogonal representation of $G$. We assume that
\begin{equation}\label{eq:V-condition}
V(a) \mathcal{H}_v = \mathcal{H}_v, \quad V(a) \mathcal{H}_{v,w} = \mathcal{H}_{v,w} V(a)
\end{equation}
for all $a \in G$ and $v,w \in \mathcal{V}$. It follows that $\hat{U}(a) = \prod_{v \in \mathcal{V}} U_v(a)$ is a global unitary symmetry of $H$.

We now compute
\begin{equation}
U_u(a) H U_u^\dagger(a) = \sum_{v \in \mathcal{V}} \mathcal{H}_v^T \mathcal{S}_v + \sum_{w \in \mathcal{N}(v)} \mathcal{S}_v^T \mathcal{H}^{u,a}_{v,w} \mathcal{S}_w,
\end{equation}
where
\begin{equation}
\mathcal{H}^{u,a}_{v,w} = \begin{cases}
\mathcal{H}_{v,w} & \text{if } u \not\in \{v, w\}\\
V(a) \mathcal{H}_{v,w} & \text{if } u = v\\
V(a^{-1}) \mathcal{H}_{v,w} & \text{if } u = w.
\end{cases}
\end{equation}
This implies that a defect in this model corresponds to the left multiplication of some $\mathcal{H}_{v,w}$ by a matrix $V(a)$ for some $a \in G$. With this motivation, we define an ancilla Hilbert space for each edge $v, w$ with orthonormal basis states $\ket{a}_{v,w}$ for all $a \in G$, and we identify $\ket{a}_{v,w}$ with $\ket{a^{-1}}_{w,v}$ \footnote{This Hilbert space can be defined rigorously as the space of square-integrable functions from $G$ to $\Complex$, provided that it is possible to define a measure on $G$. For this reason, we use the term \emph{symmetry} to refer to a group $G$ that is a locally compact topological group, so that its Haar measure is well defined. When $G$ is a continuous group, the states $\ket{a}_{v,w}$ are generally delta function normalized. For example, when $G = \Real$ they are the position eigenstates of a 1D quantum particle.}. We define a matrix of operators $\tilde{V}_{v,w}$ acting on the ancilla spaces by
\begin{equation}
\tilde{V}_{v,w} \ket{a}_{v',w'} = \begin{cases}
V(a) \ket{a}_{v',w'} & \text{if } v=v', w=w'\\
V(a^{-1}) \ket{a}_{v',w'} & \text{if } v=w', w=v'\\
\ket{a}_{v',w'} & \text{otherwise}
\end{cases}
\end{equation}
and we define the defect Hamiltonian
\begin{equation}\label{eq:general-defect}
H_{\text{defect}} = \sum_{v \in \mathcal{V}} \mathcal{H}_v^T \mathcal{S}_v + \sum_{w \in \mathcal{N}(v)} \mathcal{S}_v^T \tilde{V}_{v,w} \mathcal{H}_{v,w} \mathcal{S}_w.
\end{equation}
By construction, $H_{\text{defect}}$ decomposes into sectors corresponding to definite choices of the ancilla states, and these are precisely the set of defect configurations. We identify the original Hamiltonian $H$ in \cref{eq:general-Hamiltonian} with the sector in which the ancilla states are all the identity element of $G$.

We next show that the global symmetry $\hat{U}$ of $H$ is enriched to a local symmetry of $H_{\text{defect}}$. To do so, define a unitary representation $\tilde{U}_{v,w}$ of $G$ on the ancilla spaces by
\begin{equation}
\tilde{U}_{v,w}(a) \ket{b}_{v',w'} = \begin{cases}
\ket{a b}_{v',w'} & \text{if } v=v', w=w'\\
\ket{b a^{-1}}_{v',w'} & \text{if } v=w', w=v'\\
\ket{b}_{v',w'} & \text{otherwise}
\end{cases}
\end{equation}
and note that $\tilde{U}_{v,w}(a) \tilde{V}_{v,w} \tilde{U}^\dagger_{v,w}(a) = V(a^{-1}) \tilde{V}_{v,w}$. It follows that $\hat{U}_v(a) = U_v(a) \prod_{w \in \mathcal{N}(v)} \tilde{U}_{v,w}(a)$ is a local unitary symmetry of $H_{\text{defect}}$. It is worth noting that $\prod_{v \in \mathcal{V}} \hat{U}_v(a)$ is $\hat{U}(a)$ times the unitary which conjugates each ancilla state by $a$. Therefore $\prod_{v \in \mathcal{V}} \hat{U}_v(a)$ and $\hat{U}(a)$ coincide only on sectors for which each ancilla state lies in the center of $G$.

Finally, we present the defect model construction for the TFIM as a special case. We take
\begin{equation}\label{eq:S-H-H-TFIM}
\mathcal{S}_v = \begin{pmatrix}
X_v\\ Z_v
\end{pmatrix}, \quad
\mathcal{H}_v = \begin{pmatrix}
-g \\ 0
\end{pmatrix}, \quad
\mathcal{H}_{v,w} = \begin{pmatrix}
0 & 0\\
0 & -\frac{1}{2}
\end{pmatrix}
\end{equation}
so that \cref{eq:general-Hamiltonian} matches \cref{eq:general-H-TFIM}. We take the group $G = \Z_2 = \{1, -1\}$ and note that the representations $U_v(1) = I$, $U_v(-1) = X_v$ satisfy \cref{eq:U-condition-1,eq:U-condition-2,eq:V-condition} where
\begin{equation}
V(1) = \begin{pmatrix}
1 & 0\\
0 & 1
\end{pmatrix}, \quad
V(-1) = \begin{pmatrix}
1 & 0\\
0 & -1
\end{pmatrix}.
\end{equation}
\cref{eq:general-defect} matches \cref{eq:general-H-defect}, where $\tilde{Z}_{v,w} = \tilde{V}^{2,2}_{v,w}$. Lastly, $\tilde{U}_{v,w}(1) = I$ and $\tilde{U}_{v,w}(-1) = \tilde{X}_{v,w}$.

\section{Trivial sectors of $H_{\text{defect}}$}\label{app:trivial-sectors}
Since each sector of $H_{\text{defect}}$ is defined by the states of its ancillas, we will denote the sector with global ancilla state $\bigotimes_{v,w} \ket{A_{v,w}}_{v,w}$ by the collection of group elements $A$, which must satisfy $A_{v,w} = A^{-1}_{w,v}$. We noted in \cref{app:general-defect} that $H$ can be identified with the sector $I$ where $I_{v,w} = 1 \in G$ is the identity element for all $v, w$. We will say that a sector of $H_{\text{defect}}$ is \emph{trivial} if it is equivalent to $H$ up to conjugation by products of $U_u(a)$ operators. In this section, we prove a simple characterization of trivial sectors.

Since $\hat{U}_u(a)$ is a symmetry of $H_{\text{defect}}$, we see that conjugation by $U_u(a)$ is equivalent to conjugation by $\prod_{w \in \mathcal{N}(u)} \tilde{U}_{u,w}(a^{-1})$. This conjugation maps the sector $B$ to the sector $B'$ where
\begin{equation}\label{eq:sector-mapping}
B'_{v,w} = \begin{cases}
a B_{v,w} & \text{if } v = u, w \in \mathcal{N}(u)\\
B_{v,w} a^{-1} & \text{if } w = u, v \in \mathcal{N}(u)\\
B_{v,w} &\text{otherwise}.
\end{cases}
\end{equation}
We say that a sequence of vertices $\mathfrak{p} = (v_1, v_2, \dots, v_{n-1}, v_n)$ is a path if $v_{j+1} \in \mathcal{N}(v_j)$ for $1 \leq j \leq n-1$. If additionally $v_n = v_1$ then we say that $\mathfrak{p}$ is a cycle. For any path $\mathfrak{p}$ and sector $B$ we define the directed product
\begin{equation}
\mathfrak{p}(B) = B_{v_1, v_2} B_{v_2, v_3} \dots B_{v_{n-2}, v_{n-1}} B_{v_{n-1}, v_n}.
\end{equation}
Since any cycle passes through each vertex an even number of times, it is clear that $\mathfrak{p}(B') = \mathfrak{p}(B)$ for any cycle $\mathfrak{p}$, when $B'$ is given by \cref{eq:sector-mapping}. It follows that any trivial sector $B$ must satisfy $\mathfrak{p}(B) = \mathfrak{p}(I) = 1$ for all cycles $\mathfrak{p}$.

In order to prove the converse, suppose $B$ is a sector that satisfies $\mathfrak{p}(B) = 1$ for all cycles $\mathfrak{p}$. Let the connected components of the graph be denoted $\mathcal{V}_1, \dots, \mathcal{V}_c$. In each connected component $\mathcal{V}_j$, choose a vertex $v_j \in \mathcal{V}_j$ and then label all vertices $u \in \mathcal{V}_j$ by the directed product $B(u) = \mathfrak{p}(B)$ along any path $\mathfrak{p}$ from $v_j$ to $u$. Since the directed product along any cycle is $1$, the value of $B(u)$ does not depend on the chosen path $\mathfrak{p}$. Finally, if we apply the mapping in \cref{eq:sector-mapping} with $a = B(u)$ for each vertex $u \in \mathcal{V}$, the result is the sector $I$. We conclude that $B$ is trivial if and only if $\mathfrak{p}(B) = 1$ for all cycles $\mathfrak{p}$.

\section{Ground states of $H_{\text{TFIM}}$ are ground states of $H_{\text{defect}}$}\label{app:TFIM}
In this section, we use a stoquasticity argument similar to the one in Ref. \cite{Lieb1962} to show that every ground state of $H$ is a ground state of $H_{\text{defect}}$ when $H = H_{\text{TFIM}}$ is the Hamiltonian for the TFIM on any finite graph, as given by \cref{eq:general-Hamiltonian,eq:S-H-H-TFIM}. This result implies that bootstrap bounds from $H_{\text{defect}}$ apply to the ground states of $H_{\text{TFIM}}$ on any graph, finite or infinite.

Let
\begin{equation}
U_{\text{CNOT}} = \prod_{v \in \mathcal{V}} \prod_{w \in \mathcal{N}(v)} \text{CNOT}_{v, w},
\end{equation}
where
\begin{equation}
\text{CNOT}_{v,w} = \frac{I + Z_v}{2} + \left(\frac{I - Z_v}{2}\right) \tilde{X}_{v,w}
\end{equation}
is the controlled not gate from $v$ to $w$. We now consider a unitarily transformed Hamiltonian $H'_{\text{defect}} = U_{\text{CNOT}} H_{\text{defect}} U_{\text{CNOT}}^\dagger$ which takes the form
\begin{equation}
H'_{\text{defect}} = -\frac{1}{2} \sum_{v \in \mathcal{V}} \sum_{w \in \mathcal{N}(v)} \tilde{Z}_{v,w} - g\sum_{v \in \mathcal{V}} X_v \prod_{w \in \mathcal{N}(v)} \tilde{X}_{v,w}.
\end{equation}
Since $H'_{\text{defect}}$ commutes with each $X_v$ operator for $v \in \mathcal{V}$, we can consider separately the sectors in which each $X_v$ takes a definite $1$ or $-1$ value. Let $H^{\prime x}_{\text{defect}}$ be the Hamiltonian in the sector $X_v = x_v$. We now consider the matrix representation of $H^{\prime x}_{\text{defect}}$ in the basis in which each ancilla spin is aligned in the $+x$ or $-x$ direction. In this basis, the $\tilde{X}$ terms are diagonal and the $\tilde{Z}$ terms are off-diagonal. It follows that all matrix elements are real and all off-diagonal matrix elements are non-positive. A Hamiltonian with this property is called \emph{stoquastic} \cite{Bravyi2008} and it is easy to see that any ground state of such a Hamiltonian can be chosen to have all non-negative coefficients \cite{Lieb1962,Bravyi2008}. In fact, since it is possible to move from any basis state to any other by applying a sequence of $\tilde{Z}$ operators, it follows that any ground state can in fact be chosen to have all strictly positive coefficients.

These arguments apply for all values of $g$. In the case of $g = 0$, the conclusions can be seen explicitly. In this case, there is a unique ground state, and this state has each spin aligned in the $+z$ direction. In our basis of choice, this state has an equal positive coefficient on every basis state. In fact, since no two finite dimensional vectors with positive coefficients can be orthogonal, it follows that $H^{\prime x}_{\text{defect}}$ always has a unique ground state. Furthermore, the ground states at different values of $g$ must have the same quantum numbers for each unitary symmetry operator. For any cycle $\mathfrak{p} = (v_1,v_2, \dots, v_{n-1}, v_1)$, let
\begin{equation}
\tilde{Z}_{\mathfrak{p}} = \tilde{Z}_{v_1, v_2} \tilde{Z}_{v_2, v_3} \dots \tilde{Z}_{v_{n-2}, v_{n-1}} \tilde{Z}_{v_{n-1}, v_1}.
\end{equation}
We note that each $\tilde{Z}_{\mathfrak{p}}$ operator is a symmetry of $H^{\prime x}_{\text{defect}}$ and the ground state at $g = 0$ has eigenvalue $1$. It follows that for any value $g$, the ground state is stabilized by each $\tilde{Z}_{\mathfrak{p}}$ operator. Since $U_{\text{CNOT}} \tilde{Z}_{\mathfrak{p}} U_{\text{CNOT}}^\dagger = \tilde{Z}_{\mathfrak{p}}$, it follows that any ground state of $H_{\text{defect}}$ is stabilized by each $\tilde{Z}_{\mathfrak{p}}$ operator. Finally, the results of \cref{app:trivial-sectors} imply that the ground states of $H_{\text{defect}}$ are precisely the images of the ground states of $H_{\text{TFIM}}$ under arbitrary products of $X$ operators. In particular, every ground state of $H_{\text{TFIM}}$ is a ground state of $H_{\text{defect}}$.
\end{document}